\documentclass[amsmath,amssymb,twocolumn]{revtex4}

\usepackage{graphicx}
\usepackage{dcolumn}
\usepackage{bm}

\def\q1{{q^{-1}}}
\def\qq1{{q-q^{-1}}}

\def\nq{{n_{i}}}
\def\be{\begin{equation}}
\def\ee{\end{equation}}
\def\bee{\begin{eqnarray}}
\def\eee{\end{eqnarray}}

\begin{document}

\title{Deformed quantum statistics in two-dimensions}
\author{A. Lavagno $^{1,2}$ and P. Narayana Swamy $^3$}
\affiliation{ $^1$ Dipartimento di Fisica, Politecnico di Torino,
I-10129, Italy}
\affiliation{$^2$Istituto Nazionale di Fisica
Nucleare (INFN), Sezione di Torino, I-10126, Italy}
\affiliation{$^3$ Physics Department, Southern Illinois University,
Edwardsville, IL 62026, USA}

\begin{abstract}
It is known from the early work of May in 1964 that ideal Bose gas
do not exhibit condensation phenomenon in two dimensions. On the
other hand, it is also known that the thermostatistics arising from
$q$-deformed oscillator algebra has no connection with the spatial
dimensions of the system.
Our recent work concerns the study of important thermodynamic
functions such as the entropy, occupation number, internal energy
and specific heat in ordinary three spatial dimensions, where we
established that such thermostatistics is developed by consistently
replacing the ordinary thermodynamic derivatives by the Jackson
derivatives. The thermostatistics of $q$-deformed bosons and
fermions in two spatial dimensions is an unresolved question and
that is the subject of this investigation. We study the principal
thermodynamic functions of both bosons and fermions in the two
dimensional $q$-deformed formalism and we find that, different from
the standard case, the specific heat of $q$-boson and $q$-fermion
ideal gas, at fixed temperature and number of particle, are no
longer identical.
\end{abstract}
\maketitle

\section{Introduction}

The phenomenon of Bose-Einstein condensation of the standard Bose
gas is well established. The condensation signifies a
macroscopically large occupation number of the zero energy state
corresponding to a temperature of 3.14 K. The experimentally
observed critical temperature corresponds to 2.2 K for liquid He$^4$
when it becomes a superfluid. Indeed the phenomenon of Bose
condensation involves the macroscopically large occupation number
$N_0$ corresponding to zero energy while the average occupation
number $N_p$ corresponding to non-zero energies remains finite. The
thermostatistics of an ideal, standard,  Bose gas is well
established  and the standard behavior of all the thermodynamic
functions below and above the critical temperature is well described
\cite{Pathria}. In particular, the specific heat shows a
discontinuous behavior as a function of temperature, the point of
discontinuity generally known as the lambda point.

 It is well-known from the work of May \cite{may} that the ideal gas
obeying the standard Bose-Einstein statistics does not condense in
two dimensional space i.e., two space dimensions and one time
dimension. It is indeed known that the specific heat for an ideal
gas of Fermi particles is identically the same as that for an
ideal Bose gas for all $T$ and $N$ in two dimensions, despite the
great difference in the distribution functions of the two systems
at low temperatures. The early work of May also includes an
investigation of the extreme relativistic limit of such systems.

The question of what is implied by the absence of Bose
condensation in two dimensions has been discussed in
Ref.\cite{leepre97} and \cite{Pathria2}. This last work raises the
question of what is special about the dimensionality two. Let us
note that the above properties are rigorously true for ideal
uniform two dimensional  Fermi and Bose gases in the thermodynamic
limit \cite{swarup}. It is relevant to note that in
Ref.\cite{vafek} anomalous thermodynamics of Coulomb interacting
massless Dirac fermions in two-spatial dimensions has been
outlined and in Ref.\cite{Pilati et al}, has been studied the
thermodynamics of the interacting Bose gas in two and three
dimensions.

On the other hand, because of complicated topology of the
configuration space for indistinguishable particles in two
dimensions, Feynman's path-integral formulation allows exotic
quantum statistics which interpolates between fermions and bosons
\cite{leinaas,wu}. Analogously, quon algebra introduces $q$-deformed
commutation relation with violations of Pauli exclusion principle
and Bose statistics not related to the spatial dimension of the
system \cite{greenberg}.

In describing complex systems, quantum algebra and quantum groups
have been the subject of intensive research in several physical
fields such as cosmic strings and black holes \cite{strominger},
conformal quantum mechanics \cite{youm}, nuclear and high energy
physics \cite{bona,svira,pla02}, fractional quantum Hall effect and
high-$T_c$ superconductors \cite{wil}. From the seminal work of
Biedenharn \cite{bie} and Macfarlane \cite{mac}, it was clear that
the $q$-calculus, originally introduced by Heine \cite{heine} and by
Jackson \cite{jack} in the study of the basic hypergeometric series
\cite{gasper}, plays a central role in the representation of the
quantum groups with a deep physical meaning \cite{cele1,fink,abe}.
Furthermore, it is remarkable to observe that the $q$-calculus is
very well suited for to describe fractal and multifractal systems.
As soon as the system exhibits a discrete-scale invariance, the
natural tool is provided by Jackson $q$-derivative and $q$-integral,
which constitute the natural generalization of the regular
derivative and integral for discretely self-similar systems
\cite{erzan}.

In this framework, the thermostatistics of such deformed bosons
and fermions and the properties of $q$-deformed quantum mechanics
have been studied in Ref.\cite{pre2000,pre2002,jpa2008}. It was
found that there are consequences of the deformation in the
thermostatistics of $q$-bosons and $q$-fermions and the theory is
based on the introduction of basic numbers and it was further
shown that the thermostatistics involving the various
thermodynamic functions can be fully described if the ordinary
thermodynamic derivatives are replaced by the Jackson derivatives
\cite{jack} in a systematic manner. On general grounds, one might
remark that while the ideal gases are described by the standard
Bose-Einstein and Fermi-Dirac statistics, the statistics of real,
complex gases can thus be described by the thermostatistics based
on the $q$-deformed algebra.

In view of the above facts, it is worthwhile to ask the question:
does the $q$-boson system exhibit the phenomenon of condensation in
two space dimensions? One may ask the further question whether the
specific heat of spinless $q$-bosons and $q$-fermions, at fixed $T$
and $N$, are identical in two space dimensions. This is the precise
goal of the present work.

In this context it is relevant to observe that the Bose-Einstein
condensation of a relativistic $q$-deformed Bose gas has been
studied in Ref.\cite{Su et al} within the  standard thermodynamics
framework, not based on $q$-deformed Jackson thermodynamics
derivatives and $q$-integral. Furthermore, in Ref.\cite{potter1}
thermodynamics of ideal and statistically interacting quantum gas in
$D$ dimensions has been studied in the framework of fractional
statistics \cite{halda}.

We begin with a review of the standard quantum thermostatistics of
bosons and fermions, i.e., the undeformed gas in Section 2. The
basis of the $q$-deformed quantum thermostatistics in the framework
of $q$-calculus is presented in Section 3. The specific heat of
$q$-bosons and $q$-fermions is investigated in Section 4. The last
section contains a summary and conclusions.

\section{Undeformed quantum statistics, analytical and numerical approach}
Let us briefly review some basic properties of undeformed ($q=1$)
non-interacting bosons and fermions in two dimensions. Such results
will be very useful in the $q$-deformation extension described in
the next sections.

The grand canonical partition function for an ideal Bose gas
($\Omega_+$, $\kappa=+1$) or an ideal Fermi gas ($\Omega_-$,
$\kappa=-1$) at the temperature $T$ is given by \be
\Omega_\kappa=- \kappa\,T\, \ln {\cal Z_\kappa} \, , \ee where we
have set the Boltzmann constant equal to unity and the logarithm
of the grand partition function ${\cal Z_\kappa}$ is given by
\begin{equation}
\log {\cal Z_\kappa}=-\kappa \sum_i \ln (1-\kappa \, z \,
e^{-\beta\epsilon_i}) \; , \label{part}
\end{equation}
and $\beta=1/T$.

The average occupation number of bosons or fermions can be derived
from the relation
\begin{equation}
N_\kappa=z_\kappa \; \frac{\partial}{\partial z_\kappa} \ln {\cal
Z_\kappa}= \sum_i \frac{1}{z_\kappa^{-1}\, \exp(u_i)-\kappa} \; ,
\label{num}
\end{equation}
where $z_\kappa = e^{\beta \mu_\kappa}$ is the fugacity,
$u_i=\beta\epsilon_i$ and $\mu_\kappa$ is the chemical potential
associated with  the boson ($\kappa=+1$) or fermion ($\kappa=-1$).
For Bose statistics, the fugacity must satisfy $z_+<1$ ($\mu_+<0$,
negative chemical potential) in order to assure the non-negativity
of the  occupation numbers.

For a large (two dimensional) volume $V_2$ and a large number of
particles, the sum over all single particle energy states can be
transformed to an integral over the energy, according to
\begin{equation}
\sum_i f(u_i) \ \ \Longrightarrow \ \ \frac{V_2}{\lambda^2}
\int_0^\infty \!\!\! \, f(u)\, du \, , \label{int2d}
\end{equation}
where $u=\beta\epsilon$, $\epsilon = \hbar^2 k^2/2m$ is the kinetic
energy and $\lambda = h/(2\pi m T)^{1/2}$ is the thermal wavelength.

In the thermodynamic limit, when both $N$ and $V$ tend to infinity
but the ratio $N/V$ remains finite, the average number of
particles in Eq.(\ref{num}) can thus be written as
\begin{equation}
N_\kappa=  \frac{V_2}{\lambda^2} \int_0^\infty \!\!\! \,
\frac{1}{z_\kappa^{-1}\,\exp(u)-\kappa} \, du\; . \label{num_int}
\end{equation}
The above integral can be evaluated analytically and we obtain
\begin{equation}
N_\kappa=  -\kappa\, \frac{V_2}{\lambda^2}\, \ln(1-\kappa\,z_\kappa)
\, . \label{num2}
\end{equation}

In the case of bosons ($\kappa=1$), the right hand side of the above
equation has no upper bound and diverges logarithmically as
$z_+\rightarrow 1$, there is no temperature below which the ground
state can be said to be macroscopically occupied in comparison to
the excited states. Therefore, as it is well-known, no Bose
condensation occurs in two dimensional non-interacting Bose systems
\cite{may}. Furthermore, as first established by May \cite{may}, the
internal energies of two systems of spinless bosons and fermions at
the same fixed temperature $T$ and number of particle $N$ differ
only by a quantity proportional to $N$ and do not depend on $T$,
therefore, the two systems have the same specific heat $C_v(T,N)$.

Since we shall be exploring the same properties for $q$-deformed
bosons and fermions, let us briefly review the crucial points of
the demonstration of the above property.

The internal energy can be derived from the grand partition
function by means of the following thermodynamic derivative
\begin{equation}
U_\kappa(T,z_\kappa)=-\left. \frac{\partial}{\partial\beta} \log
{\cal Z} \right |_z= \frac{V_2}{\lambda\,\beta}\,\int_0^\infty
\!\!\! \, \frac{u}{z_\kappa^{-1}\,\exp(u)-\kappa} \, du\; .
\label{int}
\end{equation}
By introducing the Bose-Einstein and Fermi-Dirac functions \be
h_n^\kappa(z_\kappa)=\frac{1}{\Gamma(n)}\,\int_0^\infty \!\!\! \,
\frac{u^{n-1}}{z_\kappa^{-1}\,\exp(u)-\kappa} \, du \equiv
\sum_{i=1}^\infty \frac{(\kappa\, z_\kappa)^i}{i^n} \,
,\label{hn1}\ee the form for the internal energy can be cast into
the more compact expression \be U_\kappa(T,z_\kappa)=
\frac{V_2}{\lambda\,\beta}\,h_2^\kappa(z_\kappa) \, .\ee

Let us observe that the above internal energy is calculated at a
fixed temperature and fugacity or, equivalently, at a fixed number
of particles. In fact, by inverting Eq.(\ref{num2}), we have the
fugacities $z_\kappa$ as a function of $N_\kappa$ \be
z_\kappa=\kappa\,
\left[1-\exp\left(-\kappa\,\frac{\lambda^2}{V_2}\,N_\kappa\right)\right]
\, .\label{fugacity}\ee
In order to compare the internal energy at
the same $T$ and $N$, the fugacities of bosons ($\kappa=+1$) and
fermions ($\kappa=-1$) must be related by the following relations
\bee
&&z_+=1-\sigma_{_N} \, ,\\
&&z_-=\frac{1}{\sigma_{_N}}-1 \, , \eee where we have defined
$\sigma_{_N}=\exp(-N\lambda^2/V)$. The above equations can be
equivalently set as \be z_-=\frac{z_+}{1-z_+} \, . \ee

By using the property of the dilogarithmic functions \cite{may,dilo}
\be h_2^-(z_-)-h_2^+(z_+)=\frac{1}{2}\,(\ln \sigma_{_N})^2 \, ,
\label{deltah}\ee it follows that \be
U_-(T,N)-U_+(T,N)=\frac{1}{2}\, N\, \rho_{_N} \, , \label{deltau}\ee
where we have set \be
\rho_{_N}=\frac{N}{V_2}\,\frac{\hbar^2}{2\,\pi\, m} \, .\ee
Therefore, the right hand side of Eq.(\ref{deltau}) does not
explicitly depend on $T$ and the specific heat
\begin{equation}
C_v=\left. \frac{\partial U}{\partial T}\right|_{V,N}\; ,\label{cvt}
\end{equation}
at the same temperature and number of particle, are identical for
fermions and bosons \cite{may}.

As we will see in the next Section, in the $q$-deformed theory of
bosons and fermions it is not possible to find an analytical
expression analogous to  Eq.(\ref{num2}).  Eq.(\ref{fugacity}) is no
longer correct, consequently, it is crucial to test, for the
following developments, that the above properties (\ref{deltah}) and
(\ref{deltau}) can be easily obtained numerically. For further
developments, it is useful to introduce the variable $y$, defined as
\be y=\frac{N}{V_2}\,\lambda^2 \, . \label{yvar} \ee With the above
definition, the corresponding Eq.(\ref{fugacity}), for
bosons/fermions systems at the same $T$, $N$ and $V$, can be derived
as \be h_1^\kappa(z_\kappa)=y \,\,\,\Longrightarrow
\,\,\,z_\kappa=[h_1^\kappa(y)]^{-1} \, , \label{znum}\ee where the
inverse function introduced above refers to the symbolic form. We
can evaluate the internal energy as \be U_\kappa(T,N)=\frac{T\,
N}{y}\, h_2^\kappa[z_\kappa(y)]\, . \ee

In order to show that the difference of internal energy of
fermions-bosons, at the same $T$ and $N$, does not depend on $T$, it
is sufficient to show that the following equation holds \be \Delta
h_2(y)=h_2^-[z_-(y)]-h_2^+[z_+(y)]=\alpha_1\,\, y^2 \, ,
\label{deltah_num}\ee where $z_-(y)$ and $z_+(y)$ are obtained from
Eq.(\ref{znum}) and $\alpha_1$ is a dimensionless constant. In fact,
if the last equivalence of Eq.(\ref{deltah_num}) is verified, we
have \be \Delta U(N)=U^--U^+=\alpha_1\,N\,\rho_{_N}  \, .
\label{deltaU_num}\ee

Performing the numerical evaluation of the function $\Delta
h_2(y)$, we can see that it has a parabolic behavior on the
variable $y$ with the dimensionless coefficient $\alpha_1=1/2$ (as
we know analytically from Eq.(\ref{deltah})). The statistical
variable $\chi^2=\sum_i(\Delta h_2(y_i)-\alpha_1\,y_i^2)^2\approx
10^{-8}$, therefore, this numerical approach gives a very reliable
test and it will be applied in the next Section on the framework
of $q$-deformed bosons and fermions.

\section{$q$-deformed quantum statistics}

Let us briefly review the basic properties of $q$-oscillator
algebra and the generalized thermodynamic properties of
$q$-deformed bosons and fermions \cite{pre2000,pre2002}.

The symmetric $q$-oscillator algebra is defined, in terms of the
creation and annihilation operators $c$, $c^\dag$ and the
$q$-number operator $N$, by \cite{ng,chai,lee,song}
\begin{equation}
[c,c]_{\kappa}=[c^\dag,c^\dag]_{\kappa}=0 \; , \ \ \ cc^\dag- \kappa
q c^\dag c =q^{-N} \; ,
\end{equation}
\begin{equation}
[N,c^\dag]= c^\dag \; , \ \ \ [N,c]=-c\; ,
\end{equation}
where the  deformation parameter $q$ is real and $[x,\, y]_{\kappa}=
x y - \kappa y x\,$, where, as before,  $\kappa = 1$ for  $q$-bosons
with commutators and $\kappa = -1$ for $q$-fermions with
anticommutators.

Furthermore, the operators obey  the relations
\begin{equation}
c^{\dag}c = [N]\, , \ \ \ \ c c^{\dag} = [1 + \kappa N] \, ,
\label{cc}
\end{equation}
where the $q$-basic number is defined as
\begin{equation}
[x]=\frac{q^x-q^{-x}}{\qq1}\; . \label{bn}
\end{equation}

The transformation from Fock space to the configuration space
(Bargmann holomorphic representation) may be accomplished by means
the Jackson derivative (JD) \cite{jack}
\begin{equation}
{\cal D}^{(q)}_x f(x)=\frac{f(qx)-f(\q1 x)}{x\,(\qq1)}\; ,
\end{equation}
which reduces to the ordinary derivative when $q$ goes to unity.
Therefore, the JD occurs naturally in $q$-deformed structures and we
will see that it plays a crucial role in the $q$-generalization of
the thermodynamics relations.

Thermal average of an observable can be computed by following the
usual prescription of quantum mechanics. Accordingly, the
Hamiltonian of the non-interacting $q$-deformed oscillators
(fermions or bosons) expected to have the form
\begin{equation}
H=\sum_i (\epsilon_i-\mu) \, N_i\; . \label{ha}
\end{equation}
Let us note that the Hamiltonian is deformed and depends on $q$
since the number operator is deformed by means Eq.(\ref{cc}) and
it is  not linear in $c^\dag c$. Therefore, although the logarithm
of the grand partition function has the same functional expression
as in the undeformed case, Eq.(\ref{part}), the standard
thermodynamic relations in the usual form are ruled out (for
instance, it is verified that $N\ne z \, \frac{\partial}{\partial
z} \log {\cal Z}$) \cite{pre2000,pre2002}.

In Ref.\cite{pre2000}, we have shown that the entire structure of
thermodynamics is preserved if ordinary derivatives are replaced
by the use of an appropriate Jackson derivative
\begin{equation}
\frac{\partial}{\partial z} \Longrightarrow {\cal D}^{(q)}_z \; .
\end{equation}
Consequently, the number of particles in the $q$-deformed theory can
be derived from the relation
\begin{equation}
N=z \; {\cal D}^{(q)}_z \log {\cal Z}\equiv \sum_i n_i \; ,
\label{numq}
\end{equation}
where $n_i$ is the mean occupation number expressed as
\begin{equation}
n_{i}=\frac{1}{\qq1} \log\left
(\frac{z^{-1}e^{\beta\epsilon_i}-\kappa \, q^{-\kappa}} {
z^{-1}e^{\beta\epsilon_i}-\kappa \, q^\kappa}\right) \; .\label{nqi}
\end{equation}

In this context, it is relevant to observe that the statistical
origin of such $q$-deformation lies in the modification, relative
to the standard case, of number of states $W$ of the system
corresponding to the set of occupational number ${n_i}$
\cite{pre2000}. In literature, other statistical generalization
are present, such as the so-called nonextensive thermostatistics
or superstatistics with a completely different origin
\cite{tsallis,abe2}.

The usual Leibniz chain rule is ruled out for the JD and therefore
derivatives encountered in thermodynamics must be modified as
follows. First we observe that the JD applies only with respect to
the variable in the exponential form such as $z=e^{\beta \mu}$ or
$y_i=e^{-\beta \epsilon_i}$. Therefore for the $q$-deformed case,
any thermodynamic derivative of functions which depend on $z$ or
$y_i$ must be transformed to derivatives in one of these variables
by using the ordinary chain rule and then evaluating the JD with
respect to the exponential variable.  For instance, in the  case of
the internal energy in the $q$-deformed case, we can write this
prescription explicitly as
\begin{equation}
U=-\left. \frac{\partial}{\partial\beta} \log {\cal Z} \right |_z=
\kappa\sum_i \frac{\partial y_i}{\partial\beta} \, {\cal
D}^{(q)}_{y_i}\log(1-\kappa z\,y_i) \; . \label{intq}
\end{equation}
In this case we obtain the correct form of the internal energy
\begin{equation}
U=\sum_i \epsilon_i \, \nq\; , \label{un}
\end{equation}
where $n_i$ is the mean occupation number expressed in
Eq.(\ref{nqi}).

In the thermodynamic limit, for a large (two dimensional) volume
$V_2$ and a large number of particles, the sum over states can be
replaced by the integral, similar to the correspondence in
Eq.(\ref{int2d}). However, in a $q$-deformed theory the standard
integral must be consistently generalized to the $q$-integral,
inverse operator of the JD, defined, for $0<q<1$ in the interval
$[0,a]$, as \cite{exton,erzan} \be \int_0^a f(x)\, d_q
x=a\,(\q1-q)\, \sum_{n=0}^\infty q^{2n+1}\, f(q^{2n+1}\,a) \, , \ee
while in the interval $[0,\infty)$ \be \int_0^\infty f(x)\, d_q
x=(\q1-q)\, \sum_{n=-\infty}^\infty q^{2n+1}\, f(q^{2n+1}) \, . \ee

Following the above prescriptions (see, for example,
Ref.\cite{exton} for a detailed description of the $q$-integral
properties), we gain the $q$-analogue of Eq.(\ref{int2d}) as follows
\cite{wachter}
\begin{equation}
\sum_i f(u_i) \, \Longrightarrow \,I_q=\frac{V_2}{(2\pi)^2} \,
\int\!\!\! \, f[u(k)] \, d_qk_x \,\, d_qk_y \, , \label{intk}
\end{equation}
where $u(k)=\beta\, \hbar^2 k^2/2m$ and holds the constraint:
$k^2=k_x^2+k_y^2$. By taking into account  the rules related to
changing the variable of $q$-integration \cite{exton}, we have
verified that for $0.6<q<1.4$ the above integration can be well
approximately expressed as
\begin{equation}
I_q \approx \frac{V_2}{\lambda^2} \, \frac{2}{q+\q1}\, \int_0^\infty
\!\!\! \, f(u)\, d_Qu \, , \label{int2dq}
\end{equation}
where $Q=q^2$ (a change of variable $u=\beta\, \hbar^2 k^2/2m$ also
involves a corresponding change of base). Therefore, in the
thermodynamic limit, Eq.(\ref{numq}) and Eq.(\ref{intq}),
respectively, becomes
\bee &&N_{\kappa}(T,z_\kappa)=\frac{V_2}{\lambda^2} \, h_1^\kappa(z_\kappa,q) \, ,\label{nq}\\
&&U_{\kappa}(T,z_\kappa)=\frac{V_2}{\beta\,\lambda^2} \,
h_2^\kappa(z_\kappa,q)\, , \label{uq}\eee where we have defined the
$q$-deformed $h_n^\kappa(z_\kappa,q)$ as
\begin{equation}
h_n^\kappa(z_\kappa,q)=\frac{1}{\Gamma (n)} \int_0^\infty \!\!\!
 \frac{u^{n-1}}{\qq1} \log\left (\,\frac{z_\kappa^{-1}e^u-\kappa \,
q^{-\kappa}} { z_\kappa^{-1}e^u-\kappa \, q^\kappa} \right) \, d_Qu
\; . \label{hn}
\end{equation}

It must be stressed that the above equation is quite different
from the definition of the generalized function introduced in
Eq.(21) in our earlier work \cite{pre2002}. This is an important
notion in our present work. It should also be noted that, to the
best of our knowledge, this is the first time that $q$-integrals
are numerically employed in thermostatistics calculations. In the
limit $q\rightarrow 1$, the deformed $h_n^\kappa(z_\kappa,q)$
functions reduce to the standard $h_n^\kappa(z_\kappa)$ for bosons
and fermions, defined in Eq.(\ref{hn1}).

As in the undeformed boson case, we need to set the range of the
$q$-boson fugacity $z_B=z_+$ which will correspond to non-negative
occupation number. In the case of $q$-bosons we see that the
condition is $z_B< 1/q$ for $q>1$ and $z_B<1$ for $q<1$. Also in
this case the number of particle, expressed by Eq.(\ref{nq}),
diverges logarithmically as $z_B\rightarrow 1/q$ (if $q>1$) and
$z_B\rightarrow q$ (if $q<1$). Therefore, no Bose condensation
occurs in two dimensional $q$-boson gas.

Moreover, it should be pointed out that we also have to require the
existence of the JD of the mean occupation number which is
encountered in the calculation of thermodynamic quantities such as
the specific heat and this changes the upper bound of the fugacity
$z_B$. In the following, we thus will require the condition $z_B<
z_q$, where we have defined
\begin{equation}
z_q=\begin{cases} q^{-2} & \text{if $q>1$ ;} \\  q^2 & \text{if
$q<1$ .}
\end{cases}
\label{zq}
\end{equation}

At this point, we are able to see if the difference of the
internal energy of $N$ $q$-fermions and $q$-bosons at fixed $T$
does not depend on $T$ and the specific heats are equal, as in
undeformed case. These properties must be verified numerically
because of we are  unable to get the analytic expression of
Eq.(\ref{nq}), therefore, we follow the numerical procedure tested
for $q=1$ in the second part of Section II.

In the $q$-deformed theory, the previous definition of the
variable $y$ of Eq.(\ref{yvar}) must be changed with \be
y_q=\frac{q+\q1}{2}\,\frac{N}{V_2}\,\lambda^2 \, , \ee
consequently, we can obtain the fugacities
$z_\kappa=z_\kappa(y_q,q)$ as \be
z_\kappa=[h_1^\kappa(y_q,q)]^{-1} \, , \label{znumq}\ee and the
internal energy as \be U_\kappa(T,N)=\frac{T\, N}{y_q}\,
h_2^\kappa[z_\kappa(y_q,q)]\, . \ee

As before, the difference between the internal energy of fermions
and bosons  at the same $N$ and $T$, does not depend on $T$ if the
following relation holds \be \Delta
h_2(y_q,q)=h_2^-[z_-(y_q,q)]-h_2^+[z_+(y_q,q)]=\alpha_q\,\, y_q^2 \,
, \label{deltah_numq}\ee where $z_-(y_q,q)$ and $z_+(y_q)$ are
obtained from Eq.(\ref{znumq}) and $\alpha_q$ is a dimensionless
constant. It may be noted that $\alpha_q\rightarrow 1/2$ in the
limit $q\rightarrow 1$.

In Fig. 1, we plot the coefficient $\alpha_q$ for different values
of $q$, while in Fig. 2 it is possible to check the reliability of
the quadratic approximation of Eq.(\ref{deltah_numq}),
$\chi^2=\sum_i(\Delta h_2(y_{q_i},q)-\alpha_q\,y_{q_i}^2)^2$,
related to the difference of the internal energy of fermions and
bosons system at fixed $T$ and $N$. As we can see from Fig. 2, the
quadratic behavior of $h_2(y_q,q)$ holds only for small
$q$-deformation effect ($q\approx 1$) or at small value of the
variable $y$ (or the fugacity $z$), therefore, the difference of the
internal energy is not rigorously independent of the temperature and
the specific heats of bosons and fermions, at fixed $T$ and $N$ are
not exactly equal. In the next Section, we will give an explicit
evaluation of the specific heat for different values of the
deformation parameter $q$.

\begin{figure}[ht]
\centerline{\includegraphics[width=0.4\textwidth]{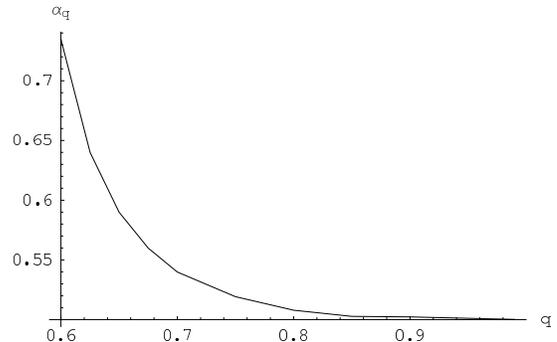}}
\caption{Plot of the dimensionless coefficient $\alpha_q$ of
Eq.(\ref{deltah_numq}) as a function of $q$.}
\end{figure}

\begin{figure}[ht]
\centerline{\includegraphics[width=0.4\textwidth]{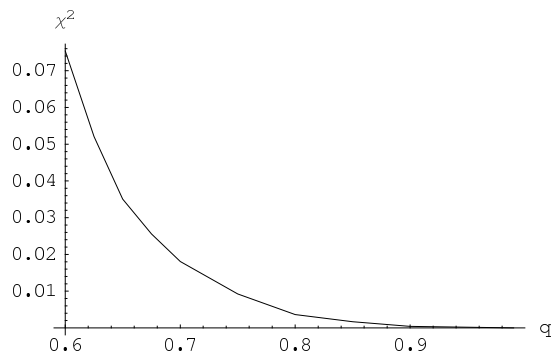}}
\caption{Behavior of the $\chi^2$ related to quadratic fit in
Eq.(\ref{deltah_numq}) as a function of $q$.}
\end{figure}

\section{Specific heat of boson and fermion systems}

We are now able to calculate the specific heat of the $q$-boson
and $q$-fermion gas, starting from the thermodynamic definition of
Eq.(\ref{cvt}).

Carrying out the JD prescription, described earlier in Sec. III,
Eq.(\ref{cvt}) in the $q$-deformed theory can be written as
\begin{equation}
C_v=-\beta^2 \sum_i \, \epsilon_i\,\frac{\partial \alpha_i}{\partial
\beta}\; \frac{1}{q-\q1} {\cal D}_{\alpha_i}^{(q)}\log \left (
\frac{1-\kappa\,
q^{-\kappa}\,\alpha_i}{1-\kappa\,q^\kappa\,\alpha_i}  \right ) \; ,
\label{cvq}
\end{equation}
where $\alpha_i= z \, e^{- \beta \epsilon_i}$ and
\begin{equation}
\frac{\partial \alpha_i}{\partial \beta}=\left(
\frac{1}{z}\frac{\partial
z}{\partial\beta}-\epsilon_i\right)\,\alpha_i\, .
\end{equation}

For this purpose we first need, therefore, the derivative of the
fugacity with respect to $T$ (or $\beta$), keeping $V$ and $N$
constant.  Accordingly, we observe that the following identity holds
(since the number of particles is kept constant)
\begin{equation}
\frac{\partial}{\partial \beta}\sum_i \log \left ( \frac{1-\kappa\,
q^{-\kappa}\,\alpha_i}{1-\kappa\,q^\kappa\,\alpha_i}  \right ) = 0
\; .
\end{equation}
In accordance with  the JD recipe about the thermodynamical
relations, the above equation can be written as
\begin{equation}
\sum_i \, \frac{\partial \alpha_i}{\partial \beta}\, {\cal
D}_{\alpha_i}^{(q)}\log \left ( \frac{1-\kappa\,
q^{-\kappa}\,\alpha_i}{1-\kappa\,q^\kappa\,\alpha_i}  \right )) =
0\; .
\end{equation}
Evaluating in the thermodynamical limit ($V\rightarrow\infty$) and
by using the definition in Eq.(\ref{hn}), we obtain
\begin{equation}
\left. \frac{1}{z}\, \frac{\partial z}{\partial\beta}\right|_{V,N}=
\frac{1}{\beta} \; \frac{{\cal D}^{(q)}_z h_2^\kappa (z,q)}{ {\cal
D}^{(q)}_z h_{1}^\kappa (z,q)} \; . \label{dzb}
\end{equation}

By using the above relation in Eq.(\ref{cvq}), we obtain the
specific heat for a system of bosons and fermions at fixed $T$ and
$N$
\begin{equation}
\frac{C_v \, \lambda^2}{V_2}\equiv \frac{C_v}{N}\,y= 2 z_\kappa\,
{\cal D}^{(q)}_z h_3^\kappa (z_\kappa,q) -z_\kappa\, \frac{({\cal
D}^{(q)}_z h_2^\kappa (z_\kappa,q))^2}{ {\cal D}^{(q)}_z h_1^\kappa
(z_\kappa,q)} \; . \label{cvfinal}
\end{equation}

In Figs. 3 and 4, we plot the behavior of the specific heat
$C_v\,\lambda^2/V_2$ for a boson and fermion system at fixed
temperature and number of particles (let us remember that we have
not taken into account the degeneracy factor due to the spin quantum
number) for two different values of $q$. As we can see from the
figures, the specific heats of boson and fermion are no longer equal
for $q\neq 1$ and this difference becomes more relevant by
increasing the value of the deformation parameter $q$.

\begin{figure}[ht]
\centerline{\includegraphics[width=0.4\textwidth]{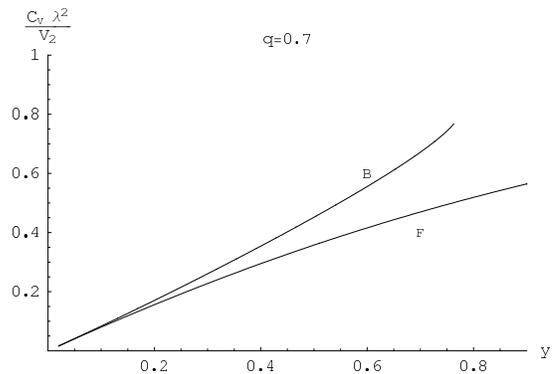}}
\caption{The specific heat $C_v\,\lambda^2/V_2$ for bosons (B) and
fermions (F), at fixed $T$ and $N$, as a function of the variable
$y$ for $q=0.7$.}
\end{figure}

\begin{figure}[ht]
\centerline{\includegraphics[width=0.4\textwidth]{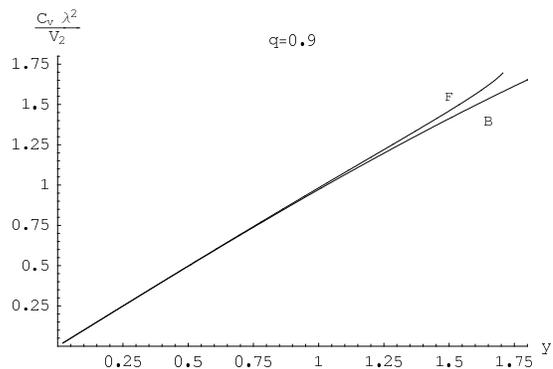}}
\caption{The specific heat as in  Fig. 3 for $q=0.9$.}
\end{figure}

It must be emphasized here that if we want to have a correct
comparison, we must plot the specific heat as a function of the
variable $y$ and not as a function of the variable
$z_{\kappa}(y_q, q)$. Same $T$ and $N$ does not imply the same
fugacity (which is a different function of $N$ and $T$ for bosons
and fermions) but the same variable $y$. To better clarify this
aspect we report in Fig. 5 and 6 the specific heat
$C_v\,\lambda^2/V_2$ as a function of the fugacity for bosons and
fermions, respectively (remember that the range of meaningful
fugacities $z_B$, for boson gas, is limited by the condition
(\ref{zq})). Let us observe that the modification of the specific
heat increasing with the value of the deformation parameter $q$
becomes very remarkable in the fermion case.

\begin{figure}[]
\centerline{\includegraphics[width=0.4\textwidth]{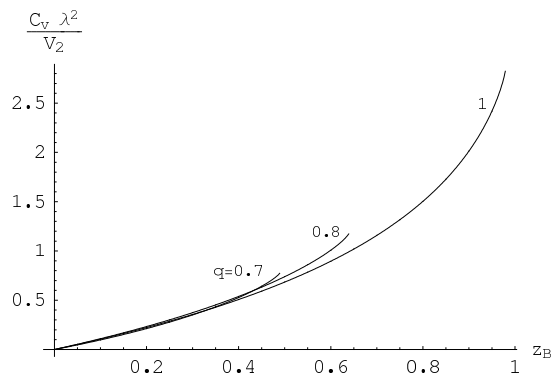}}
\caption{The specific heat $C_v\,\lambda^2/V_2$ for bosons as a
function of fugacity $z_B$ for different values of $q$.}
\end{figure}

\begin{figure}[]
\centerline{\includegraphics[width=0.4\textwidth]{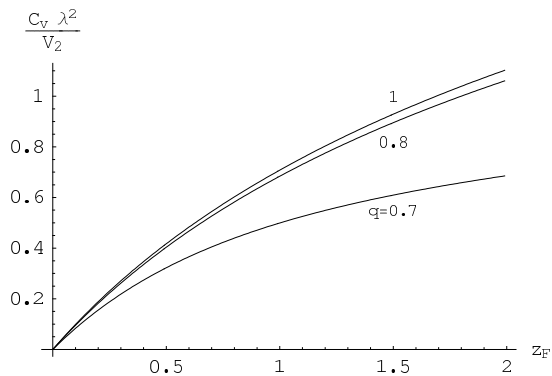}}
\caption{The specific heat $C_v\,\lambda^2/V_2$ for a fermion gas as
a function of fugacity $z_F$.}
\end{figure}

\section{Conclusion}
Understanding properties of quantum matter confined to two spatial
dimensions has been at the forefront of theoretical and
experimental physics. In the last years there was a growing
importance high energy physics, low-dimensional systems computers,
superfluid and superconducting films, quantum Hall and related
two-dimensional electron gases and low-dimensional trapped Bose
gases. On the other hand $q$-deformed quantum and statistical
theory, inspired by the quantum groups formulation, arise as the
underlying mathematical structure in several physical complex
systems.

In this paper we have investigated the structure of the
$q$-deformed quantum statistics in two-dimensions by working
consistently in the framework of the $q$-calculus with the use of
the Jackson derivatives and the $q$-integration. In this context,
we have shown that, as in the underformed case, ideal $q$-Bose gas
does not exhibit condensation, the specific heat is a continuous
function on the relevant thermodynamical variable. However, we
have shown that the difference of the internal energy of fermions
and bosons, at fixed $N$ and $T$, depends on $T$. This, as a
matter of fact, has no counterpart in the standard case and
implies that the specific heats, at fixed $N$ and $T$, of bosons
and fermions are no longer equal.

This different behavior from the undeformed quantum theory can be
dealt with in the statistical behavior of a complex systems,
intrinsically contained in $q$-deformation, whose underlying
dynamics is spanned in many-body interactions and/or long-time
memory effects. This aspect has just outlined in several papers.
For example in Ref.\cite{svira} it has been shown that
$q$-deformation plays a significant role in understanding
higher-order effects in many-body nuclear interactions. Moreover,
the strong effects on the deformation, that we have found
especially in the $q$-fermion specific heat, could be connected to
an intrinsically presence of complex many-body effective
interactions on $q$-deformation theory. In this context, it
appears relevant to observe that nonanalytic temperature behavior
of the specific heat of Fermi liquid can be explained within two
dimensional interactions beyond the weak-coupling limit
\cite{chubo}.

Let us now address a different perspective to the formulation in two
dimensions,  by adding the following remarks. The subject of anyons
has been well investigated in the recent past \cite{RAPNS}. Planar
physical systems, in two space and one time dimensions, display many
peculiar and interesting quantum properties owing to the unusual
structure of rotation, Lorentz and Poincar\'{e} groups in two
spatial dimensions and thus lead to a theory of intermediate
statistics, interpolating between Bose statistics at one end and
Fermi statistics at the other. Such anyons are described by a theory
based on the permutation group which is the braid group. Since the
real world is described  in 3+1 dimensions, anyons may not be real
particles. On one hand, the theory based only on a deformation of
the oscillator algebra which is a generalization of the ordinary
boson or fermion oscillator algebra may not have the features of a
full-fledged theory of anyons since it does not have the advantage
of the braid group characteristic of two dimensions. On the other
hand, a theory formulated on the basis of detailed balancing can
describe intermediate statistics purely on the basis of
thermostatistics and this formulation leads to a description in
terms of the basic numbers characteristic of the $q$-deformed
oscillator algebra. Conventional wisdom might indicate that such
generalization can have no relation to the algebra of deformed
harmonic oscillators since oscillators exist in any dimensions.
Accordingly, the connection between $q$-deformation and two
dimensions is an open question which has not been dealt with
satisfactorily in the literature.

\vspace{0.5cm}
\noindent {\bf Acknowledgment}\\
It is a pleasure to thank A.M. Scarfone for useful discussions.

\end{document}